\documentclass[preprint,12pt]{elsarticle}
\usepackage{xcolor}
\usepackage{subfigure}
\usepackage{epstopdf}
\usepackage{amsmath}
\usepackage{amsfonts}
\usepackage{graphicx}
\usepackage{algorithm}
\usepackage{algpseudocode}
\usepackage{amssymb}
\journal{}

\begin{document}

\begin{frontmatter}



\title{A generalized hybrid method for surfactant dynamics}


\author[mymainaddress]{Yu Fan\corref{mycorrespondingauthor}}
\cortext[mycorrespondingauthor]{Corresponding author}
\author[mymainaddress]{Shuoguo Zhang}
\author[mymainaddress]{Xiangyu Hu}
\author[mymainaddress]{Nikolaus A. Adams}
\address[mymainaddress]{Technical University of Munich Boltzmannstr. 15 D-85748 Garching }

\begin{abstract}
In this paper, we develop a generalized hybrid method for both two-dimensional (2-D) and three-dimensional (3-D) surfactant dynamics. While the Navier-Stokes equations are solved by the Eulerian method, the surfactant transport is tracked by a Lagrangian particle method, in which the remeshing technique is employed to prevent particle clustering. 
For the mass redistribution during remeshing, the redistribution weight is selected with weighted least squares, which shares the theoretical basis of the moving least squares method (MLS) and enables the present hybrid method to work in both 2-D and 3-D cases.
This optimized mass redistribution effectively strengthens the robustness of the present hybrid method, as validated by 2-D topological changes of the dumbbell.
The conservation, accuracy, and convergence of the present hybrid method have also been validated with both 2-D and 3-D test cases, including a translation circle/sphere, a deformed circle/sphere in the shear flow, and droplet deformation.
\end{abstract}


\begin{highlights}
\item A generalized hybrid method for surfactant dynamics for both 2-D and 3-D cases.
\item Mass conservation with machine precision.
\item Effectively handling topological changes without requirements of extra treatments.
\end{highlights}


\end{frontmatter}


\section{Introduction}
The presence of surfactants significantly reduces surface tension and may generate a gradient force along the interface, which conversely contributes to the mass transport on the surface. 
This coupling effect, known as surfactant dynamics, has been studied in theory \cite{gaver1990dynamics}, numerical simulation \cite{adami2010conservative,hu2007incompressible}, and experiments \cite{xu2012dynamic}. 
Among these approaches, the numerical approach, which serves as a crucial extension of theoretical and experimental research, can be primarily categorized into Eulerian, Lagrangian, and hybrid, each with its own challenges.

With Eulerian methods, the mismatch between interface and grid center generally results in mass non-conservation \cite{xu2003eulerian, adalsteinsson2003transport, xu2006level, SCHRANNER2016653, teigen2011diffuse, olshanskii2014stabilized}. In the literature, a global correction strategy \cite{SCHRANNER2016653} and separately tracking mass and volume \cite{james2004surfactant} have been proposed to resolve this issue. However, these two techniques encounter challenges in addressing local conservation and ensuring consistency in long-time simulations. 
In comparison, particle-based methods can preserve mass due to their inherent Lagrangian nature, but complication of a large number of neighbor particles significantly increases computational cost. Furthermore, with particle-based simulations, achieving a physically consistent representation of interfacial surfactants by one layer of Lagrangian particles is challenging.
A representation by multi-layer particles \cite{adami2010conservative} 
is relatively easy to implement but introduces an artificial thickness, which actually converts the 2-D surface to 3-D.
To avoid such dimensional artifacts, a co-dimension $1$ method with particle reseeding technique was proposed by Wang et al. \cite{wang2021thin}, but it does not guarantee mass conservation during reseeding. 
Hybrid methods, widely adopted in the literature, may simultaneously ensure mass conservation and enhance computational efficiency by solving fluid dynamics within the Eulerian framework while tracking surfactant evolution using either surface meshes or particles.
However, along with inheriting the advantages of Lagrangian methods, the hybrid method also inherits their drawbacks, such as mesh distortion and particle clustering. 
For these issues, various correction techniques, including artificial tangential velocity \cite{hou1994removing}, surface mapping \cite{lai2008immersed}, and surface mesh optimization \cite{botsch2004remeshing}, have been proposed, but they rely on additional topology information, i.e., the connectivity among vertices, necessitating complex data structures.
To be independent of connectivity information, Fan et al. \cite{fan2023hybrid} employed a remeshing method to handle large deformations and particle clustering. As the mass redistribution in their remeshing method shares the same theoretical foundation with Lagrange interpolation, the associated coefficient matrix becomes singular \cite{davis1975interpolation} in higher dimensions, limiting this remeshing method to 2-D problems.

In this paper, we develop a generalized hybrid method coupling an Eulerian method for fluid flow and a Lagrangian particle method for surfactant transport. The Lagrangian part incorporates a novel remeshing technique to prevent particle clustering. 
The redistribution weights during the remeshing are optimized with weighted least squares, which enables the applicability in both 2-D and 3-D, with improved robustness.
The remainder of this paper is organized as follows. Section \ref{Numerical_Preliminaries} gives a brief overview of the mathematical model and the hybrid method. 
Section \ref{mass_distribution} details the theory of the mass redistribution method and proves that the solution of mass redistribution is the same as that of MLS.
In Sections \ref{2d_case} and \ref{3d_case}, the accuracy,  convergence, and efficiency of the present method are validated with a series of 2-D and 3-D cases, respectively. 
Finally, concluding remarks are given in Section \ref{conclusion}.

\section{Numerical Preliminaries}
\label{Numerical_Preliminaries}
\subsection{Governing equations}
In a weakly-compressible two-phase fluid model, fluid flows are governed by
\begin{equation}
	\frac{\partial \mathbf{Q}}{\partial t} + \nabla \cdot \mathbf{F}
	=
	\nabla \cdot \mathbf{F}^\mu
	%\sum_{k=1}^3\frac{\partial \mathbf{F}^\mu_k}{\partial x_k}
	+\sigma \kappa \delta_\Sigma\mathbf{n}+\nabla_s \sigma\delta_\Sigma,
	\label{csv}
\end{equation}
where $\sigma$ denotes the tension coefficient on surface $\Sigma$, $ t $ the time, $ \delta_\Sigma$ the surface delta function, $\kappa$ the interfacial curvature, $\mathbf{n}$ the unit normal vector pointing outward and $\nabla_s = (\mathbf{I}-\mathbf{n}\mathbf{n})\cdot\nabla$ the surface gradient operator with $\mathbf{I}$ denoting the identity matrix. 
$\mathbf{Q}=(\rho, \rho u_1, \rho u_2, \rho u_3)$ represents conserved quantities, where $ \rho $ denotes the fluid density and $ u_{i} $ the velocity component. 
$\mathbf{F}=(\mathbf{F}_1,\mathbf{F}_2,\mathbf{F}_3)$ and $\mathbf{F}^\mu=(\mathbf{F}^\mu_1,\mathbf{F}^\mu_2,\mathbf{F}^\mu_3)$ are fluxes of conserved quantities and viscous fluxes. 
The second and third terms $ \sigma \kappa \delta_\Sigma\mathbf{n} $ and $ \nabla_s \sigma\delta_\Sigma $ on the right side of Eq. (\ref{csv}) are surface tension and Marangoni force, respectively.

To close the system of Eq. (\ref{csv}), additional equations are introduced, including the equation of state (EoS) for fluids, a constitutive relation for surfactant, a surfactant transport equation and an interface representation.

Under the weakly-compressible assumption, the EoS is written as
\begin{equation}
	p=B\left[\left(\frac{\rho}{\rho_0}\right)^\gamma-1\right]+p_0.
	\label{weakly}
\end{equation}
Here, $\gamma = 7.15$ and $ p $ the pressure. $\rho_0$ and $p_0$ are the reference density and pressure, respectively. The artificial speed of sound is given as $c^2 = \gamma(p-p_0+B)/\rho$.
To ensure near incompressible behavior, i.e., $c \geq 10\max|\mathbf{u}|$ with $\mathbf{u}$ denoting the velocity vector, the coefficient $B$ should be sufficiently large.

Following Ref. \cite{adami2010conservative}, the surface tension coefficient is estimated by
\begin{equation}
	\sigma = \max(\hat{\sigma}(1-\beta C),0),
	\label{constitutive}
\end{equation}
where $\hat{\sigma}=1.5\sigma_0$ and $\beta=1/3$.
$\sigma_0$ is the tension coefficient of clean interface, and $C$ surfactant concentration.
For a clean interface, the surfactant concentration remains unchanged, i.e., $C=1$, which leads to $\sigma = \sigma_0$.

The surfactant concentration $C$ in Eq. (\ref{constitutive}) is governed by an 
interfacial transport equation
\begin{equation}
	\frac{\partial C}{\partial t}+\nabla_{s} \cdot (C\mathbf{u})=D\nabla^2_{s}C,
	\label{concentration}
\end{equation}
where $\nabla_s^2$ is the Laplace-Beltrami operator , and $D$ the diffusion coefficient.

The interface is implicitly described by the zero level-set, i.e., $\phi = 0$, and $\phi$ is governed by
\begin{equation}
	\frac{\partial \phi}{\partial t} + \mathbf{u} \cdot \nabla \phi =0.
	\label{lsm}
\end{equation}
Here, the outward normal vector and curvature of the interface can be approximated by $\mathbf{n} = \nabla \phi/|\nabla \phi|$ and $\kappa=\nabla \cdot \mathbf{n}$, respectively. Note that, for basic validations, the velocity vector $ \mathbf{u} $ can be assigned directly or obtained from the geometric information, instead of solving the Navier-Stokes equations.

\subsection{Framework of the hybrid method}
Following the framework of the hybrid method proposed by Fan et al. \cite{fan2023hybrid},  the transport equation of surfactant in present work is solved by a Lagrangian particle method, while fluid dynamics are calculated by an Eulerian method. 
The referenced hybrid method \cite{fan2023hybrid} is briefly summarized as Algorithm \ref{hybrid}.
\begin{algorithm}
	\caption{Hybrid method for passive scalar transport}
	\begin{algorithmic}[1]
		\Procedure{\bf{: Initialization{$ \hfill\triangleright $Execute only once}}}{}
		%\State Provide the thresholds of particle quality $Q_0$ and $Q_1$.
		\State Initialize the grid data, including fluid states and level-set field. 
		\State Generate particles with proper mass on the interface. 
		\EndProcedure
		\Procedure{\bf{: Simulation iterations}}{}
		\State Estimate the surfactant concentration and surface tension on the interface.
		\State Advancement of particle position, fluid states, and level-set field.
		\Procedure{\bf{: Adaptive remeshing control}}{}
		\State Evaluate the quality $Q$ of particle distribution.
		\If{$Q < Q_0$ (uniform distribution)}
		\State Do nothing.
		\ElsIf{$Q_0<Q < Q_1$(slightly non-uniform distribution)}
		\State Particle relaxation, and mass redistribution.
		\ElsIf{$Q > Q_1$ (poor distribution)}
		\State Particle resampling, particle relaxation, and mass redistribution.
		\EndIf
		\EndProcedure
		\EndProcedure
		
	\end{algorithmic}
	\label{hybrid}
\end{algorithm}

In the initialization, particles are generated at cut-cell center, and then projected onto the interface by
\begin{equation}
	\label{projection}
	\mathbf{x} \leftarrow \left[\mathbf{x} -\phi(\mathbf{x})\mathbf{n}(\mathbf{x})\right],
\end{equation}
where $ \mathbf{x} $ represents the particle position. The particle mass with respect to particle $i$ is assigned by $m_i = C^0(\mathbf{x}_i)v_i$, with $C^0$ and $ v_i $ denoting the initial concentration and particle volume, respectively.

During the simulation iterations, the concentration $C$ and its gradient $ \nabla_s C $ are estimated by kernel approximation, with respect to particle $ i $
\begin{equation}
	\left\{
	\begin{aligned}
		C_i &= \Sigma_j{W_{ij}m_j}\\
		\nabla_s C_i &= \Sigma_j{\nabla_i W_{ij}(C_i-C_j)v_j}
	\end{aligned}
	\right.,
\end{equation}
where the subscript $j$ represents neighboring particles. Here, $W_{ij}=W(\Vert \mathbf{x}_i - \mathbf{x}_j\Vert,h)$ with $h$ and $W$ denoting the cut-off radius and the $5^{th}$-order Wendland function, respectively.
Consequently, upon estimating the concentration and its gradient, the surface tension and Marangoni force could be approximated, followed by the advancement of fluid states and level-set field \cite{hu2007incompressible}. 

During flow evolution, the quality $ Q $ of the particle distribution needs to be assessed periodically, and different strategies of particle regularization are executed
\begin{equation}
	Q = \frac{\max\{|v_i-\bar{v}|\}}{\bar{v}}.
	\label{quality}
\end{equation}
Here, the particle volume is estimated by $v_i = 1/\sum_{j} W_{ij}$, and the average particle volume is $\bar{v}=\sum_{i} v_i/N$, with $N$ denoting the total number of particles. 

In detail, when $Q_1>Q>Q_0$, the non-uniformly distributed particles can be directly replicated as a new particle cloud, and then regularized by particle relaxation \cite{zhu2021cad} with $M$ (a small integer) iterations. When $Q>Q_1$, particles of the new particle cloud should be resampled at cut-cell center and then projected onto the interface, followed by a particle relaxation with sufficient iterations to reach a uniform particle distribution ($Q<Q_0$).
After particle relaxation, the particle mass in the previous particle cloud $\{\mathbf{x}_p\}$ is redistributed to the new (present) cloud $\{\mathbf{x}_q\}$.

\section{The mass redistribution method}
\label{mass_distribution}
\subsection{Limitation of the Lagrange-based mass redistribution method}\label{original_method}
In the Lagrange-based mass redistribution method \cite{fan2023hybrid}, the surfactant concentration $ C $ at previous and present time steps are expressed as 
\begin{equation}
	\begin{cases}
		C^{prev}(\mathbf{x}) = \Sigma_p{W(\Vert \mathbf{x}-\mathbf{x}_p\Vert,h)m_p}\\[3mm]
		C^{prsnt}(\mathbf{x}) = \Sigma_q{W(\Vert \mathbf{x}-\mathbf{x}_q\Vert,h)m_q}
	\end{cases}.
\end{equation} 
Here, the subscripts $p$ and $q$ represent particles belonging to previous and present particle clouds $\{\mathbf{x}_p\}$ and $\{\mathbf{x}_q\}$, respectively.

According to the following rule of mass redistribution,
the particle mass of the present cloud can be obtained from that of the previous cloud
\begin{equation}
	m_q = \Sigma_p{m_p\beta_{p,q}},
\end{equation}
where $\beta_{p,q}$ is the weight of mass redistribution from particle $p$ to $q$.

Hence, the surfactant concentration difference $ \epsilon(\mathbf{x}) $ between $ C^{prev}(\mathbf{x}) $ and $ C^{prsnt}(\mathbf{x}) $ can be 
calculated through
\begin{equation} \label{error}
	\begin{split}
		\epsilon(\mathbf{x}) &= C^{prev}(\mathbf{x})-C^{prsnt}(\mathbf{x}) \\
		&=\Sigma_p{W(\Vert \mathbf{x}-\mathbf{x}_p\Vert,h)m_p} - \Sigma_q{W(\Vert \mathbf{x}-\mathbf{x}_q\Vert,h)m_q} \\
		&=\Sigma_p{W(\Vert \mathbf{x}-\mathbf{x}_p\Vert,h)m_p} - \Sigma_q{W(\Vert \mathbf{x}-\mathbf{x}_q\Vert,h)\Sigma_p{m_p\beta_{p,q}}}\\
		&=\Sigma_p{m_p[W(\Vert \mathbf{x}-\mathbf{x}_p\Vert,h)- \Sigma_q{W(\Vert \mathbf{x}-\mathbf{x}_q\Vert,h)\beta_{p,q}}]}
	\end{split}.
\end{equation}	

Following Ref. \cite{fan2023hybrid}, $W(\Vert \mathbf{x}-\mathbf{x}_q\Vert,h)$ can be expanded on the tangent plane with local coordinates, here to second order as
\begin{equation}
	\begin{aligned}
		W(\Vert \mathbf{x}-\mathbf{x}_q\Vert,h) &= W(\Vert \mathbf{x}-\mathbf{x}_p\Vert,h) + \sum_{i=1}^2\frac{\partial W}{\partial x^i}\bigg|_{(\mathbf{x}-\mathbf{x}_p)}(x_q^i-x_p^i) \\
		&+ \frac{1}{2}\sum_{i=1}^2\sum_{j=1}^2\frac{\partial^2 W}{\partial x^i \partial x^j}\bigg|_{(\mathbf{x}-\mathbf{x}_p)}(x_q^i-x_p^i)(x_q^j-x_p^j) + O(h^3),
	\end{aligned}
\end{equation}
where non-bold $x$ with a superscript is the local coordinate. Note that the superscripts $i,j = 1,2$ in the present subsection represents the corresponding component. 

Canceling terms until $O(h^3)$, we have following linear system with 6 equations
\begin{equation}
	\left\{
	\begin{aligned}
		&\sum_q \beta_{p,q}=1\\
		&\sum_q \beta_{p,q}(x_q^i-x_p^i)=0,\ i=1,2\\
		&\sum_q \beta_{p,q}(x_q^i-x_p^i) (x_q^j-x_p^j)=0,\ i,j=1,2
	\end{aligned}
	\right..
	\label{ls}
\end{equation}

For the 1-D interface in \cite{fan2023hybrid}, the system Eq. (\ref{ls}) degenerates into 
\begin{equation}
	\left\{
	\begin{aligned}
		&\sum_q \beta_{p,q}=1\\
		&\sum_q \beta_{p,q}(x_q-x_p)=0\\
		&\sum_q \beta_{p,q}(x_q-x_p)^2=0
	\end{aligned}
	\right..
	\label{lsd}
\end{equation}
By choosing a suitable number of neighbor particles, the reduced system Eq. (\ref{lsd}) has a unique solution because its coefficient matrix is a transformed Vandermonde matrix.
However, for higher dimensions, the unisolvence for Eq. (\ref{ls}) does not hold \cite{davis1975interpolation}, and there are infinite solutions for this linear system.
To overcome this limitation, inspired by the MLS method, we can select the optimal solution with the weighted least-squares redistribution weights, which will be detailed in Section \ref{MLS-based_improved_redistribution_method}.

\subsection{The MLS-based mass redistribution method}
\label{MLS-based_improved_redistribution_method}
The linear system Eq. (\ref{ls}) can be rewritten as
\begin{equation}
	\sum_q \beta_{p,q}P_k(\mathbf{x}_q-\mathbf{x}_p)=\delta_{0k},\ k=0,\cdots,5,
\end{equation}
where $\delta_{0k}$ is the Kronecker delta, and the function $P_k$ is defined as following monomials
\begin{equation}
	\left \{
	\begin{aligned}
		P_0 &= 1\\
		P_1 &= (x^1)\\
		P_2 &= (x^2)\\
		P_3 &= (x^1)^2\\
		P_4 &= (x^1)(x^2)\\
		P_5 &= (x^2)^2
	\end{aligned}
	\right..
	\label{constraint}
\end{equation}
Here, the superscript of $x$ within the bracket is the component index of the local coordinates of corresponding vector $\mathbf{x}$, while that of outside the bracket is the exponent.

By regarding Eq. (\ref{constraint}) as constraints, we can select the optimal solution with the weighted least-squares redistribution weights
\begin{equation}\label{MLS}
	\min S = \sum_q \eta(\Vert \mathbf{x}_q-\mathbf{x}_p \Vert,h)\beta_{p,q}^2,
\end{equation}
where the penalty function $\eta$ is defined through
\begin{align*}
	\eta(\Vert \mathbf{x}_q-\mathbf{x}_p \Vert,h)
	=\left\{
	\begin{aligned}
		&\frac{1}{W(\Vert \mathbf{x}_q-\mathbf{x}_p \Vert,h)}  \quad &0\leq \Vert \mathbf{x}_q-\mathbf{x}_p \Vert < h\\
		&\quad\quad\quad\ \infty  &else \\
	\end{aligned}
	\right..
	\label{penalty}
\end{align*}
Note that such a target function $S$ results in less negative weights with smaller magnitude, which improves the robustness of the redistribution method.

Consequently, $ \boldsymbol{\beta}_{p}=\{\beta_{p,q}\} $ has the same solution as that of MLS \cite{levin1998approximation}
\begin{equation}
	\boldsymbol{\beta}_{p} = \mathbf{D}^{-1}\mathbf{EA}^{-1}\mathbf{c}.
\end{equation}
Where $ \mathbf{A} = \mathbf{E}^T \mathbf{D}^{-1}\mathbf{E} $ with the superscript $T$ denoting the transpose operation, and $\mathbf{c} = (1, 0, ..., 0)^T$. With $m$ denoting the number of neighbor particles, matrices $\mathbf{E}$ and $\mathbf{D}^{-1}$ are defined as 
\begin{equation}
	\mathbf{E} = \left (
	\begin{array}{ccc}
		P_0(\mathbf{x}_1) &... &P_5(\mathbf{x}_1) \\
		... &... &... \\
		P_0(\mathbf{x}_m) &... &P_5(\mathbf{x}_m) \\
	\end{array}
	\right),
\end{equation}
and
\begin{equation}
	\mathbf{D}^{-1} = \left (
	\begin{array}{cccc}
		W(\Vert \mathbf{x}-\mathbf{x}_1 \Vert,h) &0 &... &0 \\
		0 &W(\Vert \mathbf{x}-\mathbf{x}_2 \Vert,h) &... &0\\
		...&...&...&...\\
		0 &... &W(\Vert \mathbf{x}-\mathbf{x}_{m-1} \Vert,h) &0 \\
		0 &... &0 &W(\Vert \mathbf{x}-\mathbf{x}_m \Vert,h) \\
	\end{array}
	\right).
\end{equation}

Note that, in the present work, the MLS method is utilized for determining the
$\beta_{p,q}$ for the approximation of $W(\Vert \mathbf{x}-\mathbf{x}_p \Vert,h)$ from $W(\Vert \mathbf{x}-\mathbf{x}_{q}\Vert,h)$,  instead of the concentration interpolation.
The details of the MLS method and other properties can be found in \cite{levin1998approximation}.

\subsection{The tangential plane and local coordinates}
As mentioned in Section \ref{original_method}, $W(\Vert \mathbf{x}-\mathbf{x}_q\Vert,h)$ is expanded on the tangent plane with local coordinates. 
In 2-D, the tangent plane is actually a line, and the sole basis vector of the tangent line can be directly computed through the normal vector of the line. However, in 3-D, two basis vectors of the tangent plane $ \mathbf{t}_1 $ and $ \mathbf{t}_2 $ have to be constructed.
In the present work, we construct the tangent plane at each particle, the normal vector $ \mathbf{n} $ is obtained by interpolation at each particle,
and the two basis tangent vectors can be obtained as follows. 

In the support domain of particle $p$ in $\{\mathbf{x}_p\}$, we select the present particle farthest away from particle $p$ , denoted as particle $q$, and then the basis vector $\mathbf{t}_1$ is defined as
\begin{equation}
	\mathbf{t}_1 = 
	\frac{\mathbf{t}_1'}{\Vert\mathbf{t}_1'\Vert},
	%\frac{\mathbf{x}_q- \mathbf{x}_p}{\Vert \mathbf{x}_q- \mathbf{x}_p \Vert}.
\end{equation}
where $\mathbf{t}_1'=\mathbf{r}_{pq}-(\mathbf{r}_{pq} \cdot \mathbf{n})\mathbf{n}$ is the projection of the vector $\mathbf{r}_{pq}=\mathbf{x}_q-\mathbf{x}_p $ onto the tangent plane.

The tangent vector $ \mathbf{t}_2 $ is obtained by the cross product of $ 
\mathbf{t}_1 $ and the normal vector $ \mathbf{n} $
\begin{equation}
	\mathbf{t}_2 = \mathbf{t}_1 \times \mathbf{n}.
\end{equation}

\section{2-D Validations}
\label{2d_case}
In this section, to investigate the effectiveness of present
remeshing method in 2-D case, all the 2-D test cases in 
Ref. \cite{fan2023hybrid}, including a translation circle, 
a deformed circle in the shear flow, topological changes 
of the dumbbell and the droplet deformation, are simulated 
here with analysis.
\subsection{A translational circle}
\label{translational_circle}
To verify the accuracy and convergence of the present remeshing 
method,
a translational circle with an initial concentration field prescribed 
on its circular interface is firstly studied here. While the translation of the circle already eliminates the 
influence from its physical deformation on the remeshing accuracy, 
other possible factors, such as the number of relaxation iterations 
and the remeshing frequency, should also be excluded to ensure a 
focused assessment of the accuracy of the present remeshing method. Consequently, the adaptive remeshing control in Algorithm \ref{hybrid} 
is not applied, and an excessive number of relaxation iterations 
are intentionally implemented for the remeshing at each time step. 

\begin{figure}[tb!]
	\centering     
	\subfigure[Initial state ($ t=0 $, left panel) and final state ($ t=1 $, right panel)]{
		\begin{minipage}{0.9\linewidth}
			\includegraphics[width=1\textwidth]{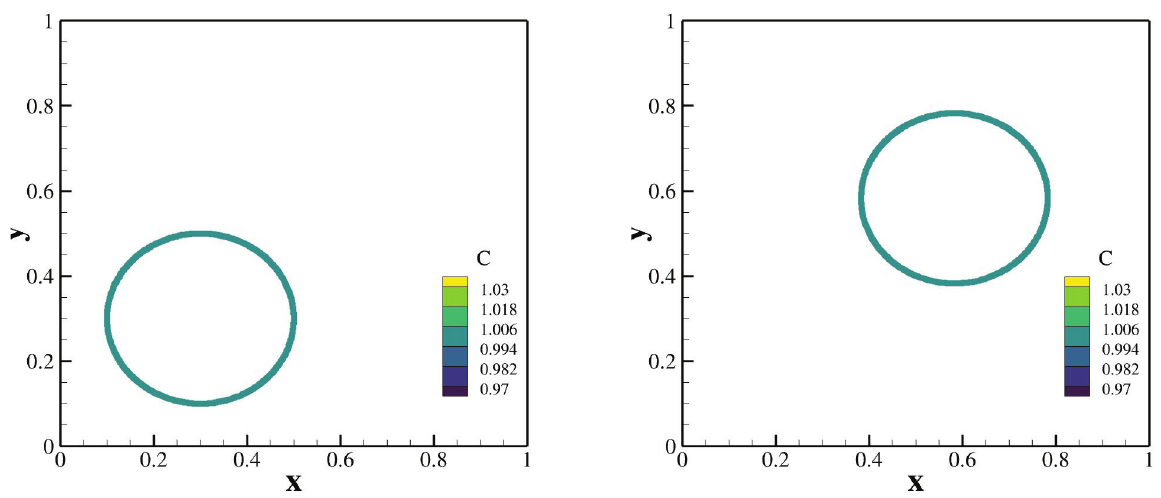}
		\end{minipage}
	}	
	\subfigure[Initial state ($ t=0 $, left panel) and final state ($ t=1 $, right panel)]{
		\begin{minipage}{0.9\linewidth}
			\includegraphics[width=1\textwidth]{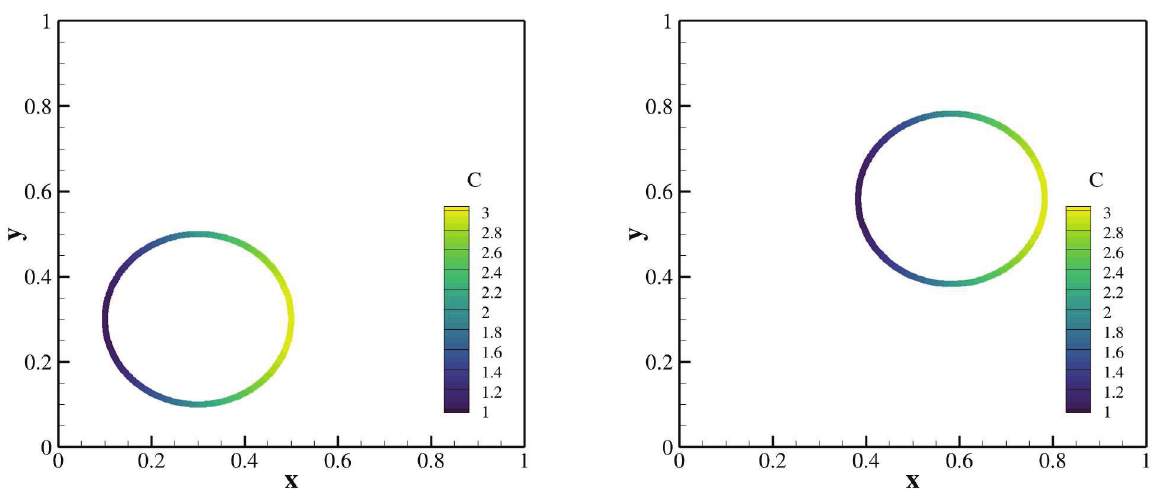}
		\end{minipage}
	}
	\caption{Evolution of initially uniform (top panel) and non-uniform (bottom panel) concentration fields during the circle's translation. The computational domain has a $256\times 256$ grid.}
	\label{Evolution_of_circle}
\end{figure}

\begin{figure}[H]%[tb!]
	\centering
	\includegraphics[width=0.6\textwidth]{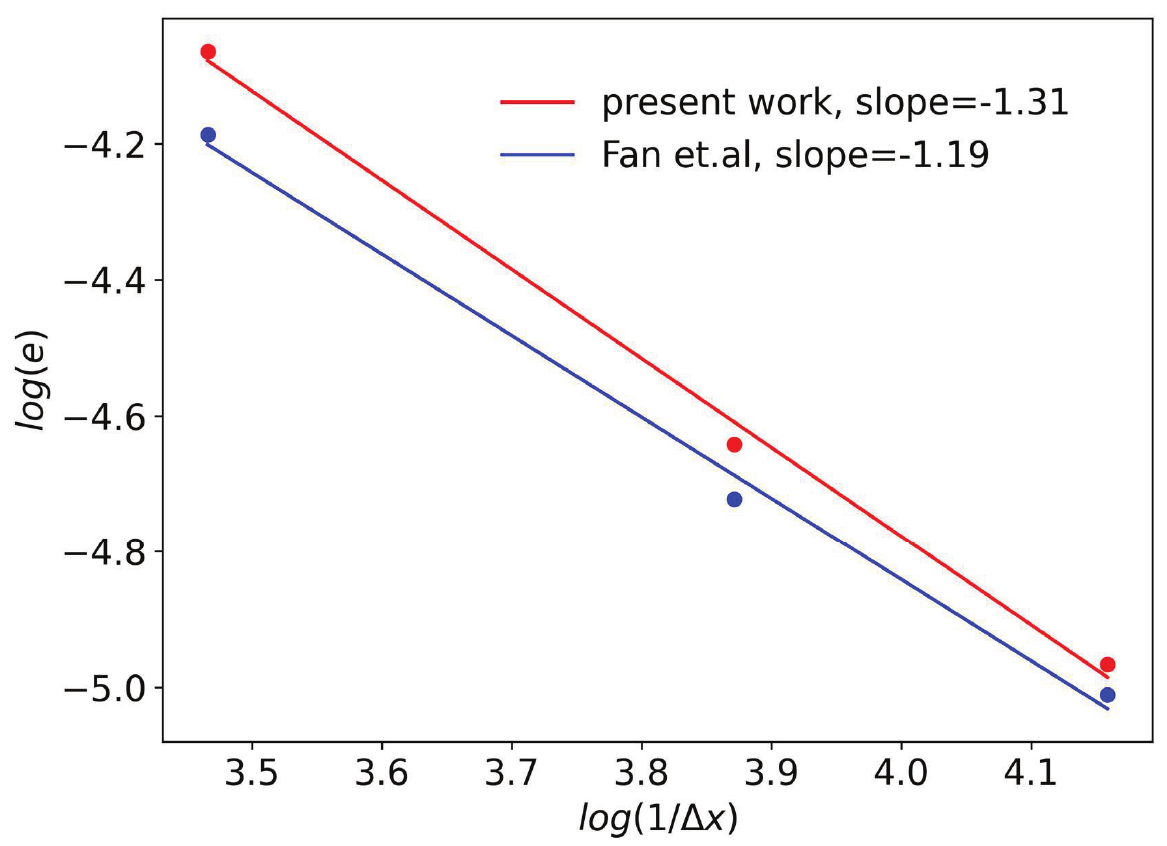} 
	\caption{Convergence analysis based on the initially non-uniform concentration field. $ \Delta x $ represents the grid size, and $ e $ the L1 error of concentration field.}
	\label{comparsion-nonuniform}
\end{figure}

In the unit-square computational 
domain, a circle with the radius of $r=0.2$ is initially centered at 
$(x_c, y_c)=(0.3, 0.3)$, and then translated with the 
velocity $(u , v) = (0.4/\sqrt{2}, 0.4/\sqrt{2}) $.
Two different initial concentration fields, i.e., a uniform one with $C(\theta)=1$ and a non-uniform one with $C(\theta)=2+\cos\theta$, are considered. Here, $ \theta $ represents the angle in polar coordinates with the center of the circle as its origin. 
Fig. \ref{Evolution_of_circle} shows the evolution of these two concentration fields.
During the circle's translation, both the initially uniform and non-uniform concentration fields are preserved well, demonstrating the accuracy of the present 
remeshing method.
In Fig. \ref{comparsion-nonuniform}, the convergence analysis is conducted on the initially non-uniform concentration field, and the 
order of convergence is approximately 1, which is close to that Fan et al. \cite{fan2023hybrid}.

\subsection{Deformed circle in shear flow}
\label{2dshear}
To further investigate the accuracy of the remeshing method when considering the geometry deformation and the adaptive remeshing control, a flexible circle with a 
radius of $r=1/3$ is placed in a shear flow. 
In the unit-square computational domain, the circle is centered at $(x_c, y_c)=(0.5, 0.5)$, and the shear flow is  
prescribed with the velocity profile $ (u, v)=(0, 0.5(x-0.5))$.
The analytical solution of the concentration field depending on time $ t $ and the angle $ \theta $ in polar coordinates is given as \cite{fan2023hybrid}  
\begin{equation}
	C(\theta,t)=\frac{1}{\sqrt{1+0.5^2t^2\sin^2\theta-t\sin\theta \cos\theta}}.
\end{equation}

\begin{figure}[H]%[tb!]
	\centering
	\includegraphics[width=0.9\textwidth]{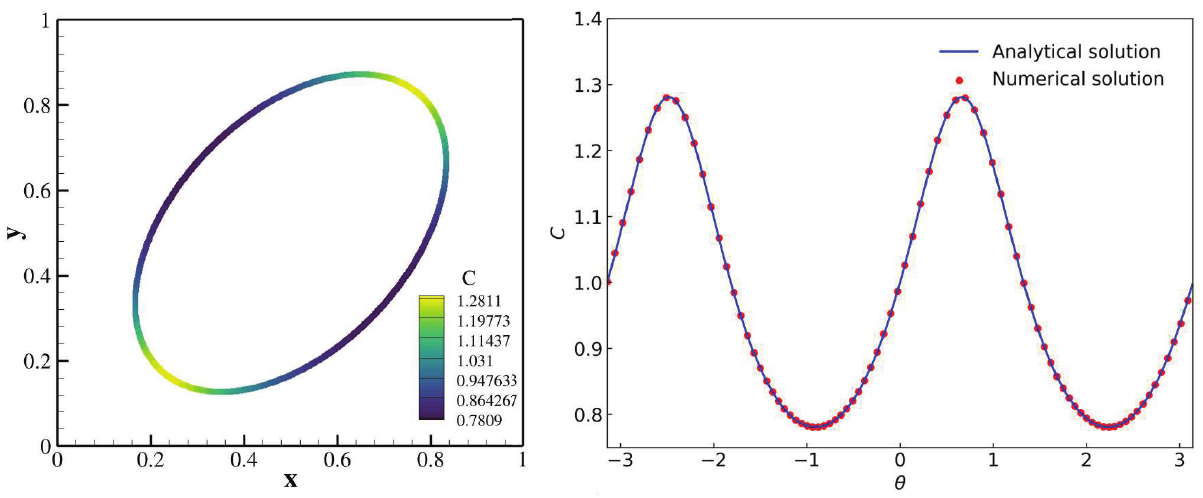} 
	\caption{The concentration distribution at $ t=1 $, and the comparison of the numerical result with analytical solution. The computational domain has a $256\times256$ grid.}
	\label{givenshear2d}
\end{figure}

\begin{figure}[H]%[tb!]
	\centering
	\includegraphics[width=0.6\textwidth]{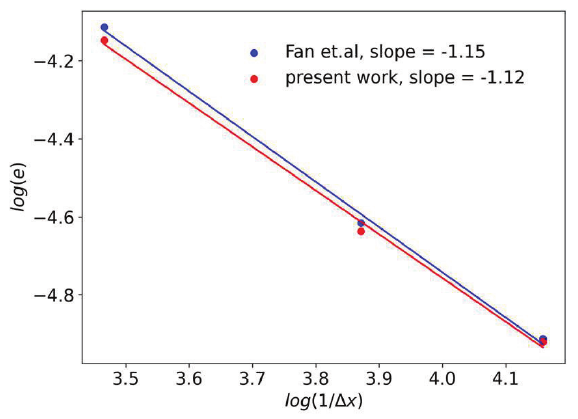} 
	\caption{Convergence analysis.}
	\label{convegence_shear}
\end{figure}

The left panel of Fig. \ref{givenshear2d} shows the concentration distribution at $ t=1 $. The upper right and lower right interfaces are stretched with a low concentration, while the tips have a higher concentration. When further compared with the corresponding analytical solution, a good agreement is demonstrated in the right panel of Fig. \ref{givenshear2d}.
Furthermore, with the same way for the convergence analysis in subsection \ref{translational_circle}, the 1st-order convergence 
is also obtained in Fig. \ref{convegence_shear}, which is close to that of Fan et al.\cite{fan2023hybrid}.

\subsection{Topological changes of a dumbbell}
\begin{figure}[H]%[tb!]
	\label{dumbbellcase}
	\centering
	\subfigure[Shape of the dumbbell]{
		\begin{minipage}[b]{0.7\textwidth}
			\includegraphics[width=1\textwidth]{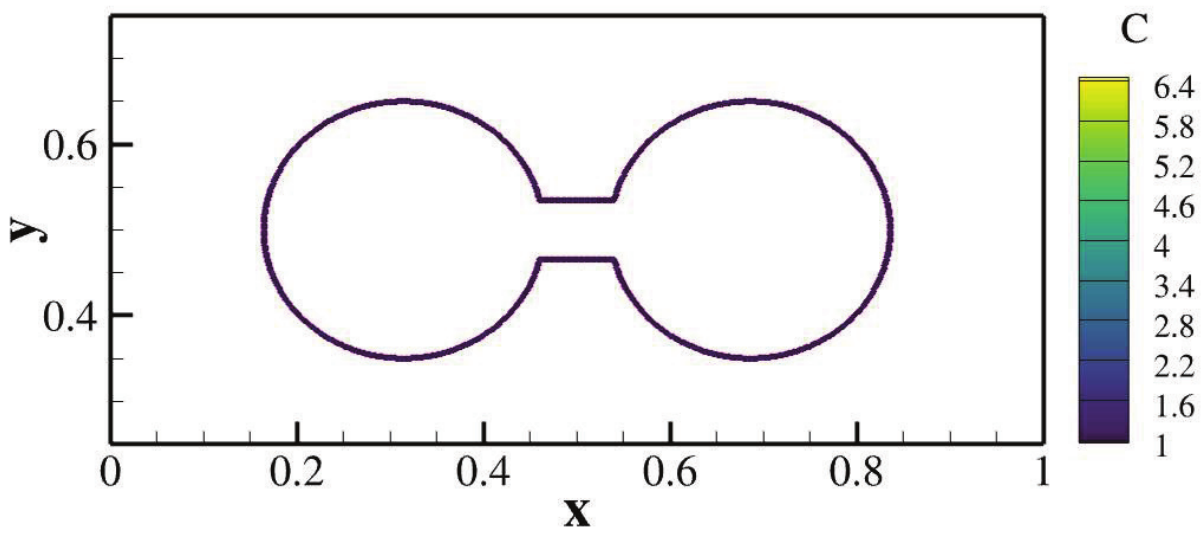}
		\end{minipage}
		\label{fig:dumbbell}
	}
	\\
	\subfigure[Before break-up]{
		\begin{minipage}[b]{0.7\textwidth}
			\includegraphics[width=1\textwidth]{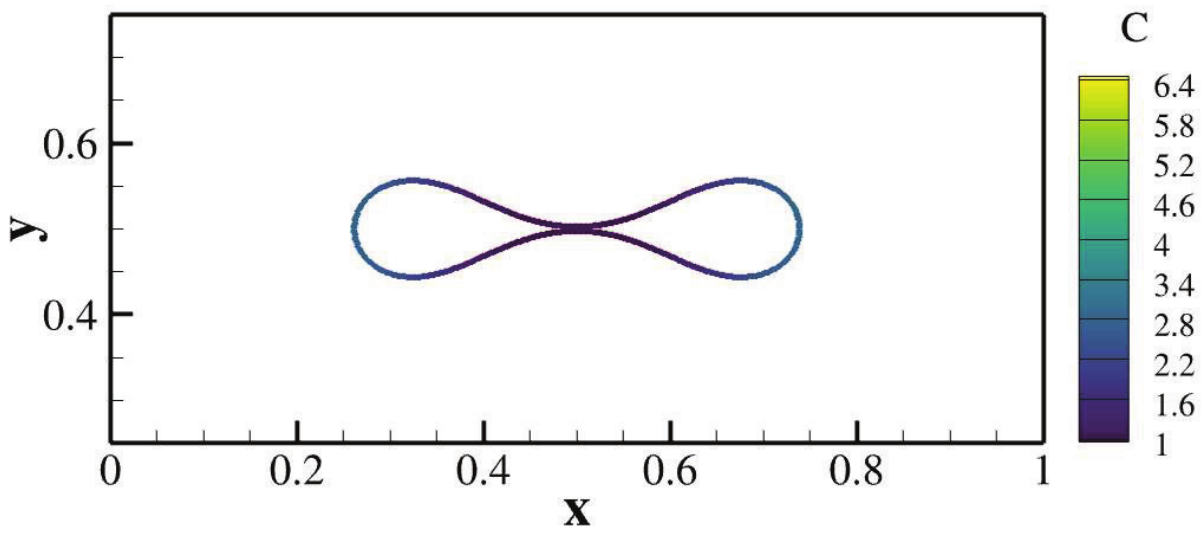}
		\end{minipage}
		\label{fig:contact}
	}
	\\
	\subfigure[After break-up]{
		\begin{minipage}[b]{0.7\textwidth}
			\includegraphics[width=1\textwidth]{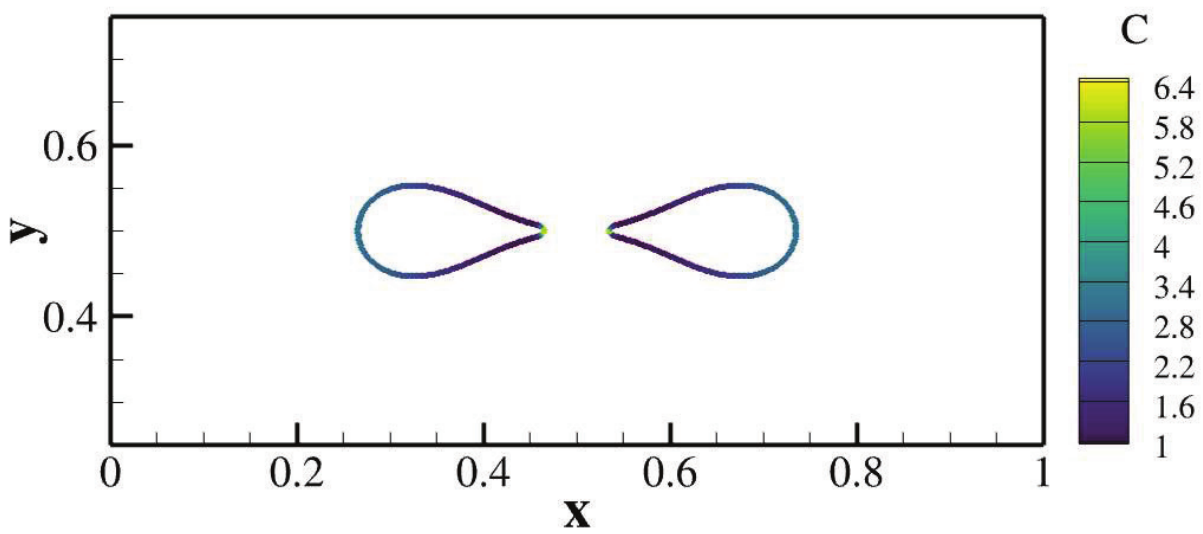}
		\end{minipage}
		\label{fig:break}
	}
	\\
	\subfigure[Final shapes]{
		\begin{minipage}[b]{0.7\textwidth}
			\includegraphics[width=1\textwidth]{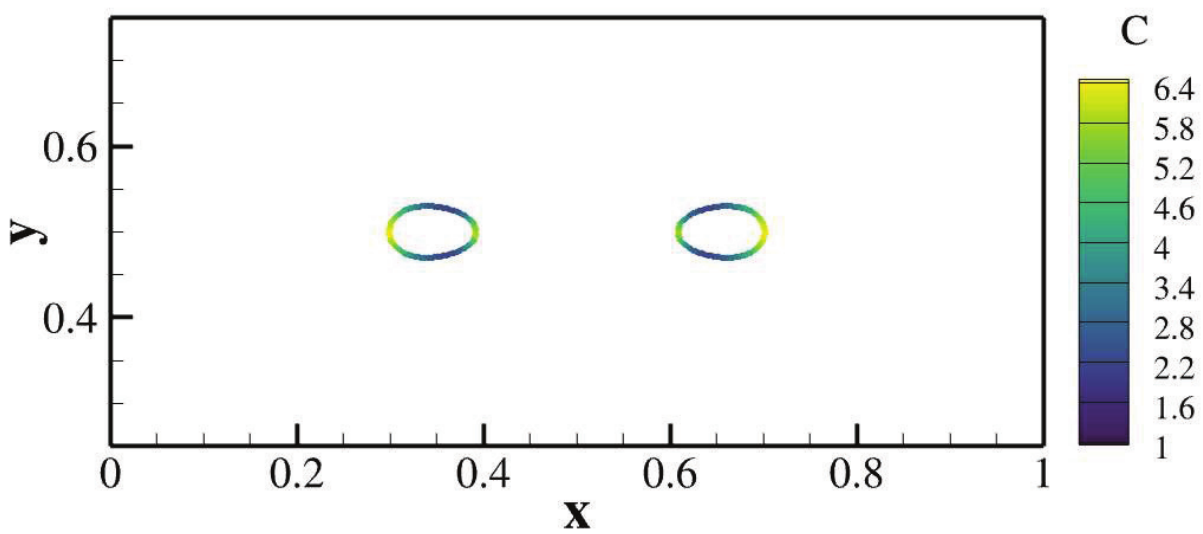}
		\end{minipage}
		\label{fig:finalshapes}
	}
	\caption{Topological changes of the dumbbell interface in a curvature driven flow with diffusion coefficient $D = 0.1$ and a $512 \times 512$ grid.}
	\label{Dumbbell}
\end{figure}

To investigate the capability of the present remeshing method in handling topological changes, a 2-D dumbbell interface is initially modeled, subsequently deformed by the curvature driven flow and ultimately torn. Fig. \ref{fig:dumbbell} shows the schematic of the dumbbell interface that the radius of both circles is $r = 0.15$ and the 
size of the rectangular shaft is $0.1\times0.07$. The velocity profile of the curvature driven flow is given as $\mathbf{u} = -0.1\kappa \mathbf{n} - 2 \mathbf{n}$. 

In Fig. \ref{fig:contact}, once the upper and lower interfaces approach each other with a distance less than the cut-off radius, 
the upper-interface particles will be identified as the neighboring particles of lower-interface particles numerically, and vice versa.
In present work, without accurately identifying neighboring particles in physics as in Ref. \cite{fan2023hybrid}, both upper- and lower-interfaces are straightforward regarded as interconnected, i.e., an 'X'-shaped junction, and the present remeshing method can still maintain a reasonable concentration field for this case by minimizing the weighted least-squares redistribution weights, i.e., Eq. (\ref{MLS}). As the time proceeds,  the dumbbell in Fig. \ref{fig:break} is torn at some point, and finally seperates into two independent circles in Fig. \ref{fig:finalshapes}.

\subsection{Droplet deformation}
To further evaluate the effectiveness and performance of the present hybrid method in surfactant dynamics, the droplet deformation in the shear flow \cite{luo2015conservative,adami2010conservative,hu2007incompressible, taylor1934formation, taylor1932viscosity} is studied with a 2-D two-phase model.
The droplet with the radius of $R_0$ is centered at $(x_c, y_c)=(4R_0,4R_0)$ in the square computational domain with the size of $8R_0 \times 8R_0$.
The droplet and the shear flow have the same density $\rho$ but different viscosity $\eta_d$ and $\eta_w$.
To realize the shear flow, two solid walls are respectively set to the top and bottom of the computational domain with opposite velocities $u=\pm u_\infty$, while the periodic boundary condition is implemented on left and right sides. Other relevant parameters are listed in Tab \ref{tab:para2d}.

\begin{figure}[H]%[tb!]
	\centering
	\subfigure[Lagrange-based redistribution \cite{fan2023hybrid},  $\mathcal{D}=0.1585$]{
		\begin{minipage}[b]{0.45\textwidth}
			\includegraphics[width=1\textwidth]{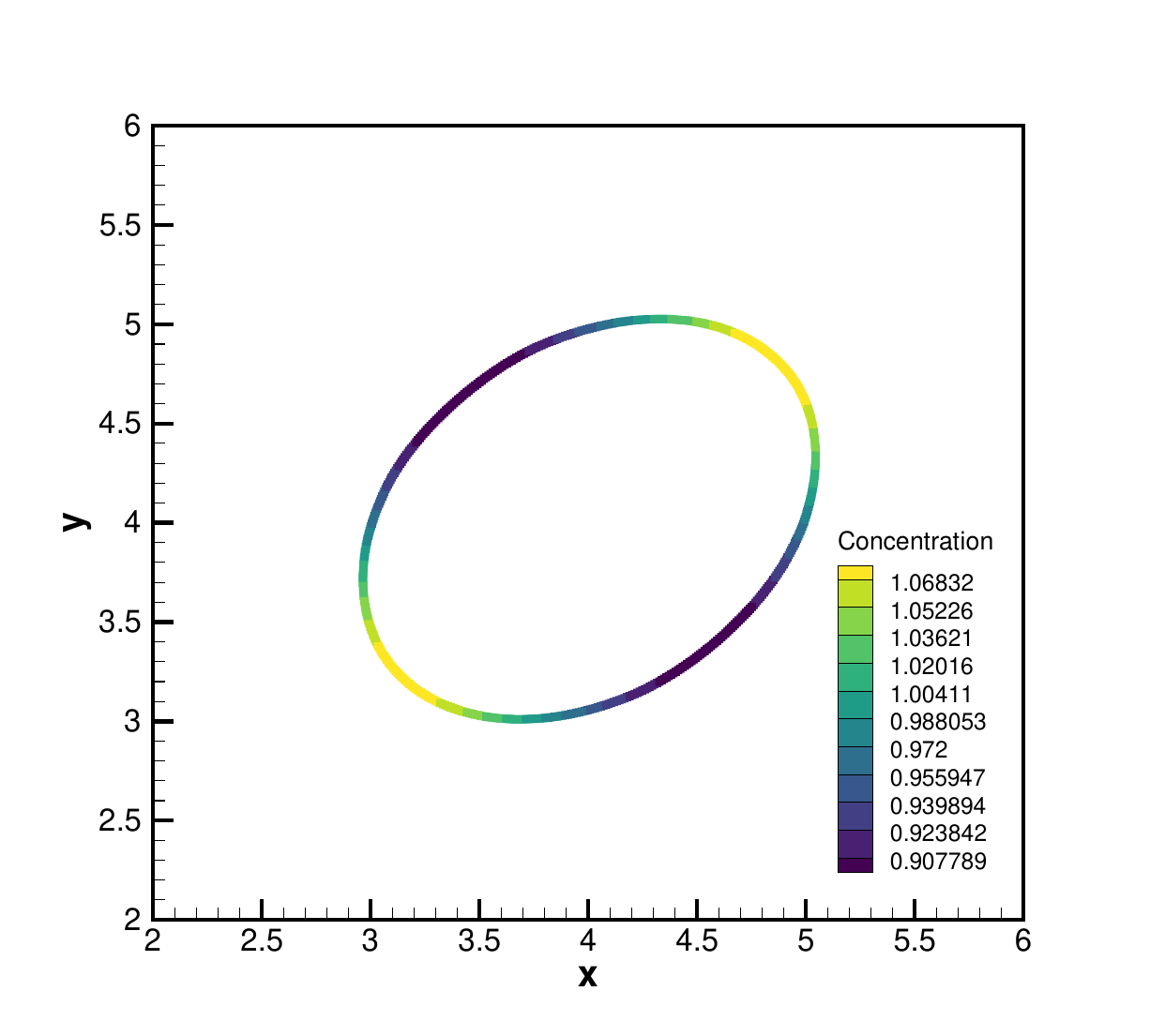} 
		\end{minipage}
	}
	\subfigure[MLS-based redistribution, $\mathcal{D}=0.1622$]{
		\begin{minipage}[b]{0.45\textwidth}
			\includegraphics[width=1\textwidth]{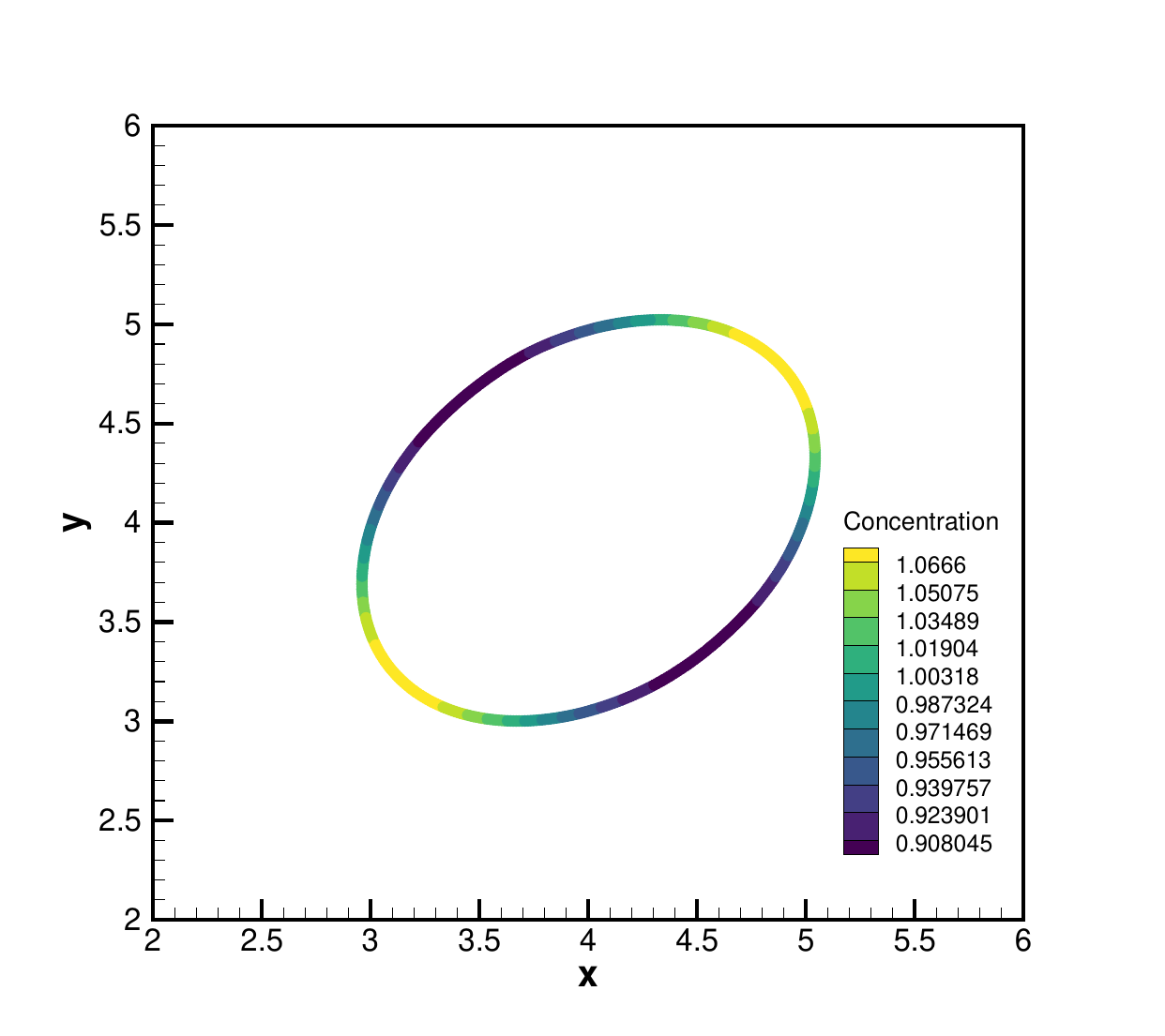} 
		\end{minipage}
	}
	\caption{The comparison of the 2-D droplet deformation with two mass redistribution methods at time instant $ t=8 $. The computational domain has a $512\times 512$ grid.}
	\label{fig:2ddroplet}
\end{figure}

\begin{figure}[H]%[tb!]
	\centering
	\includegraphics[width=0.6\textwidth]{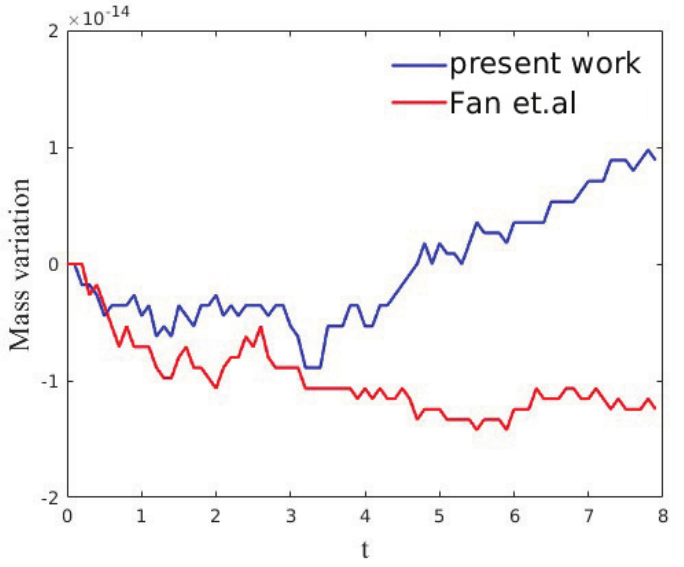} 
	\caption{Comparison of mass conservation during droplet deformation.}
	\label{conservationcompare}
\end{figure}

The deformation degree can be quantitatively expressed with a deformation parameter $\mathcal{D}=(L-B)/(L+B)$, where $L$ and $B$ are the major and minor axis, respectively.
In Fig. \ref{fig:2ddroplet}, deformation degrees of the droplet in two different mass redistribution methods are compared with small difference, which proves the effectiveness of present remeshing method in complicated flow.
Fig. \ref{conservationcompare} demonstrates an order of magnitude for the mass loss, which is comparable to that of the previous work and is attributed to machine error.
Furthermore, the ratio between the time cost of the remeshing process and the total computing time is less than 1$ \% $, which also demonstrates the good efficiency of present remeshing method.
\begin{table}[H]%[tb!]
	\centering
	\caption{Parameters for 2-D droplet deformation}
	\label{tab:para2d}
	\begin{tabular}{|c|c|c|}
		\hline
		\textbf{Parameters} &\textbf{Definitions} &\textbf{Value}  \\ \hline
		Ratio of viscosity & $\lambda=\eta_d/\eta_\omega$ & 1 \\ \hline
		Shear rate & $G=2u_\infty/8R_0$ & 1 \\ \hline
		Reynolds number & $Re=\rho G R_0^2/\eta_\omega$ & 1 \\ \hline
		Capillary number & $Ca=G\eta_\omega R_0/\sigma_0$ & 0.15 \\ \hline
		Peclet number & $Pe=GR_0^2/D$ & 1 \\ \hline
	\end{tabular}
\end{table}


\section{3-D test cases}
\label{3d_case}
To demonstrate the effectiveness of present hybrid method in higher dimensions, three 2-D benchmark cases in section \ref{2d_case}, including a translational circle, a deformed circle in the shear flow and the droplet deformation, are correspondingly modeled in three dimensions. Note that, all other parameters and settings in 2-D cases remain unchanged in the following 3-D simulations unless mentioned. 
\subsection{Sphere translation}
A sphere with the radius of $r = 0.2$ is initially centered at $(x_c, y_c,z_c)=(0.3, 0.3, 0.3)$ in the cubic computational domain with the size of $[0,1]\times [0,1] \times[0,1]$, and then translated with the velocity $ (u, v, w)=(0.4/\sqrt{3}, 0.4/\sqrt{3},0.4/\sqrt{3}) $. Grid resolution is set as  $64\times64\times64$. Here, we also consider two different initial concentration fields, i.e., a uniform one with $C=1$ and a non-uniform one with $C(x,y,z)=(z+r)/(2r)+1$.

\begin{figure}[H]%[tb!]
	\centering
	\subfigure[Initially uniform concentration field with the level-set method applied. A contour depression occurs during time evolution.]
	{
		\begin{minipage}[b]{0.8\textwidth}
			\includegraphics[width=1\textwidth]{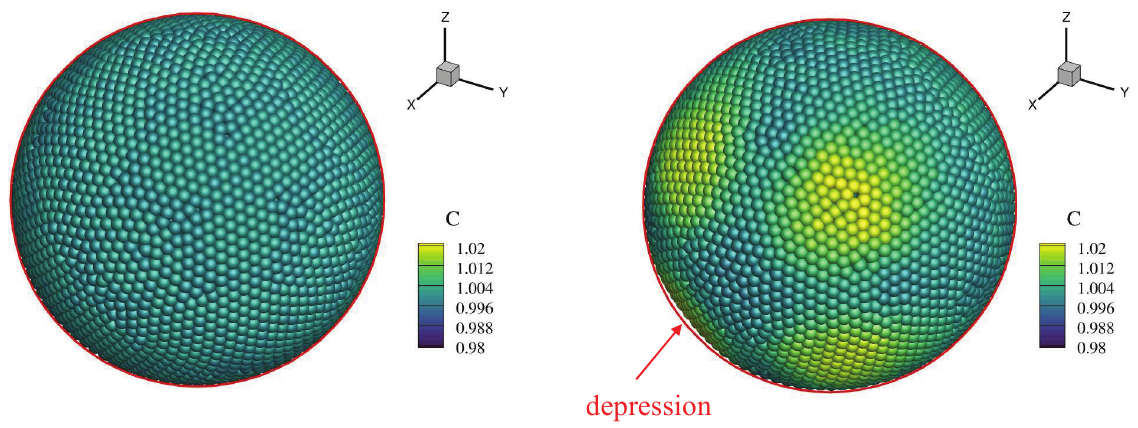} 
		\end{minipage}\label{fig:3duniform}
	}
	\subfigure[Initially uniform concentration field with the exact level-set field prescribed. No contour depression occurs during time evolution.]{
		\begin{minipage}[b]{0.8\textwidth}
			\includegraphics[width=1\textwidth]{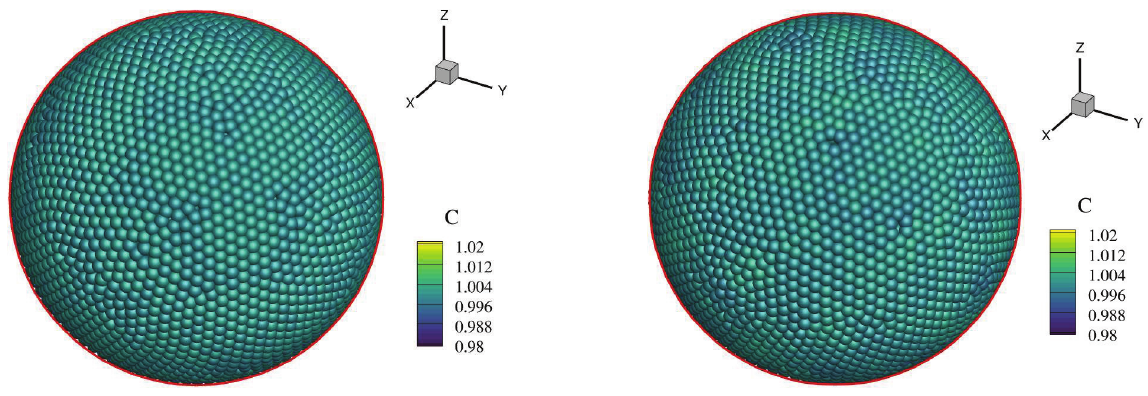} 
		\end{minipage}\label{fig:exact}
	}
	\subfigure[Initially non-uniform concentration field with the level-set method applied.]{
		\begin{minipage}[b]{0.8\textwidth}
			\includegraphics[width=1\textwidth]{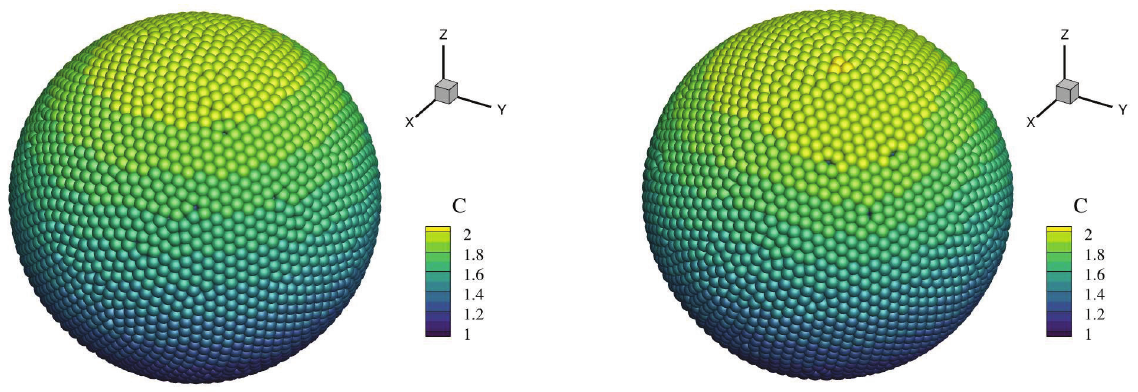} 
		\end{minipage}\label{fig:3dnonuniform}
	}
	\caption{3-D evolution of the concentration field on a sphere. The red circle represents the ideal sphere shape. Initial state ($ t=0 $, left panel) and final state ($ t=1.0 $, right panel).}
\end{figure}

For the initially uniform case, the concentration distribution becomes non-uniform, as shown in Fig. \ref{fig:3duniform}. 
This is because the level-set method can not precisely maintain the geometric shape during the translation for the given resolution, which influences the concentration evolution. 
For confirmation, an exact level-set field has also been prescribed during translation.
With the exact level-set field, we find that the geometry shape and the concentration field can be maintained well during translation, as shown in Fig. \ref{fig:exact}.
For the non-uniform case in Fig. \ref{fig:3dnonuniform}, even with the influence from the level-set method, the concentration field can be well maintained.

\subsection{Deformed sphere in shear flow}
In the unit-cube computational domain a sphere with the radius of $r=0.2$ is centered at $(x_c, y_c,z_c)=(0.5,0.5,0.5)$, and a shear flow is prescribed with the velocity profile $(u, v,w)=(0,0.5(x-0.5),0)$. 
The full derivation of the analytical solution of the concentration field is carried out in \ref{derivation}. The final result is
\begin{equation}
	C(\phi,\theta,t)=\frac{1}{\sqrt{0.5^2t^2\sin^2(\phi)\sin^2(\theta)-t\sin^2(\phi)\sin(\theta)\cos(\theta)+1}},
\end{equation}
where $(r,\theta,\phi)$ is the spherical polar coordinates and the origin is at the center of the given sphere.

\begin{figure}[H]%[tb!]
	\centering
	\includegraphics[width=0.9\textwidth]{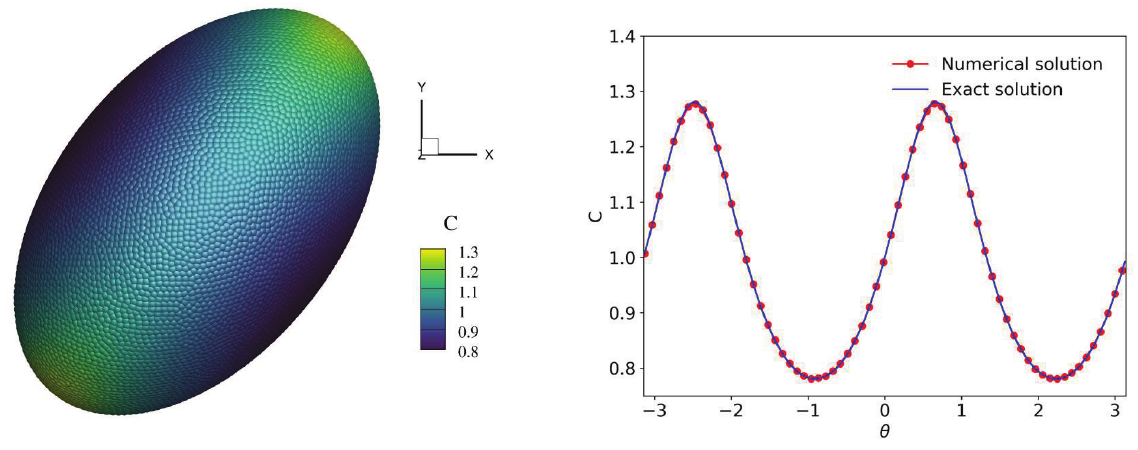} 
	\caption{The concentration distribution at $ t=1 $, and the comparison of the numerical results with analytical solution on cross-section $xOy$. The computational domain has a $128\times128\times 128$ grid.}
	\label{givenshear3d}
\end{figure}

\begin{figure}[H]%[tb!]
	\centering
	\includegraphics[width=0.6\textwidth]{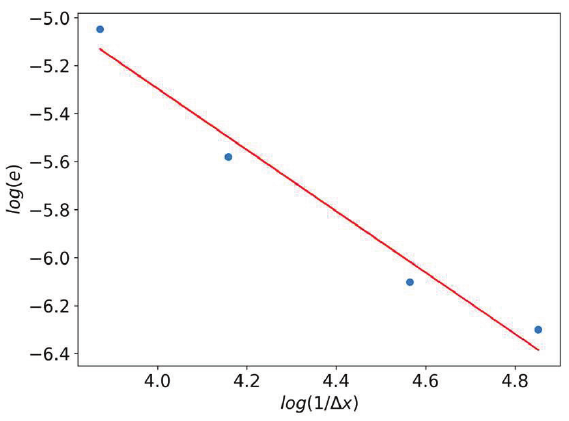} 
	\caption{Convergence analysis.}
	\label{convegence_shear3d}
\end{figure}

The left panel of Fig. \ref{givenshear3d} shows the concentration field at $t=1$. The sphere is stretched in $y$ direction and becomes an ellipsoid, with higher concentration near the two tips and lower concentration on the other parts.
We further compare the numerical concentration with the analytical one on the cross-section $xOy$, and a good agreement is shown in the right panel of Fig. \ref{givenshear3d}.
With the convergence analysis as in subsection \ref{translational_circle}, we find that it also has 1st-order convergence, as shown in Fig. 
\ref{convegence_shear3d}, which is consistent with the 2-D case in subsection \ref{2dshear}.

\subsection{Droplet deformation}
\begin{figure}[H]%[tb!]
	\centering
	\includegraphics[width=0.5\textwidth]{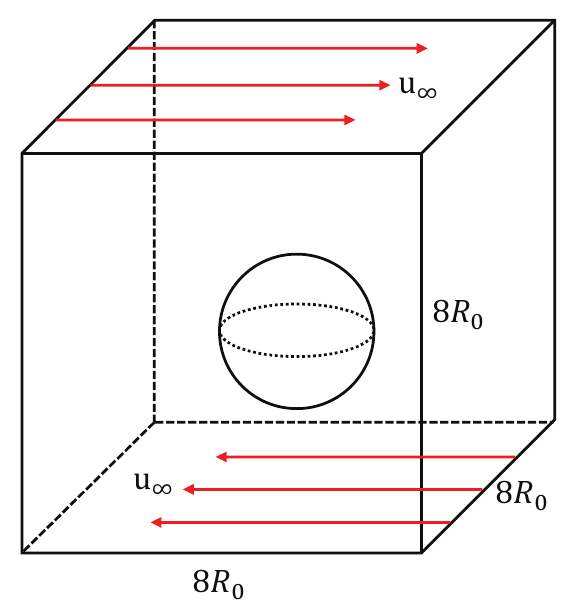}
	\caption{Schematic of 3-D droplet deformation in two-phase flow.}
	\label{3ddropconfig}
\end{figure}

As depicted in Fig. \ref{3ddropconfig}, a droplet with radius $ R_0 $ is centered at $(x_c, y_c,z_c)=(4R_0, 4R_0, 4R_0)$ in a cubic computational domain with the size of $8R_0 \times 8R_0 \times 8R_0$.
Two solid walls are respectively set to the top and bottom of the computational domain with opposite velocities $u=\pm u_\infty$, and the periodic boundary condition is implemented on other four faces. As the flow evolves, the droplet finally deforms into a ellipsoid, and the deformation degree $\mathcal{D}=(L-B)/(L+B)$ also can be calculated through the major and minor axis $ L $ and $ B $.

\begin{figure}[H]%[tb!]
	\centering
	\subfigure[Marangoni force (red arrow)]{
		\begin{minipage}[b]{0.8\textwidth}
			\includegraphics[width=1\textwidth]{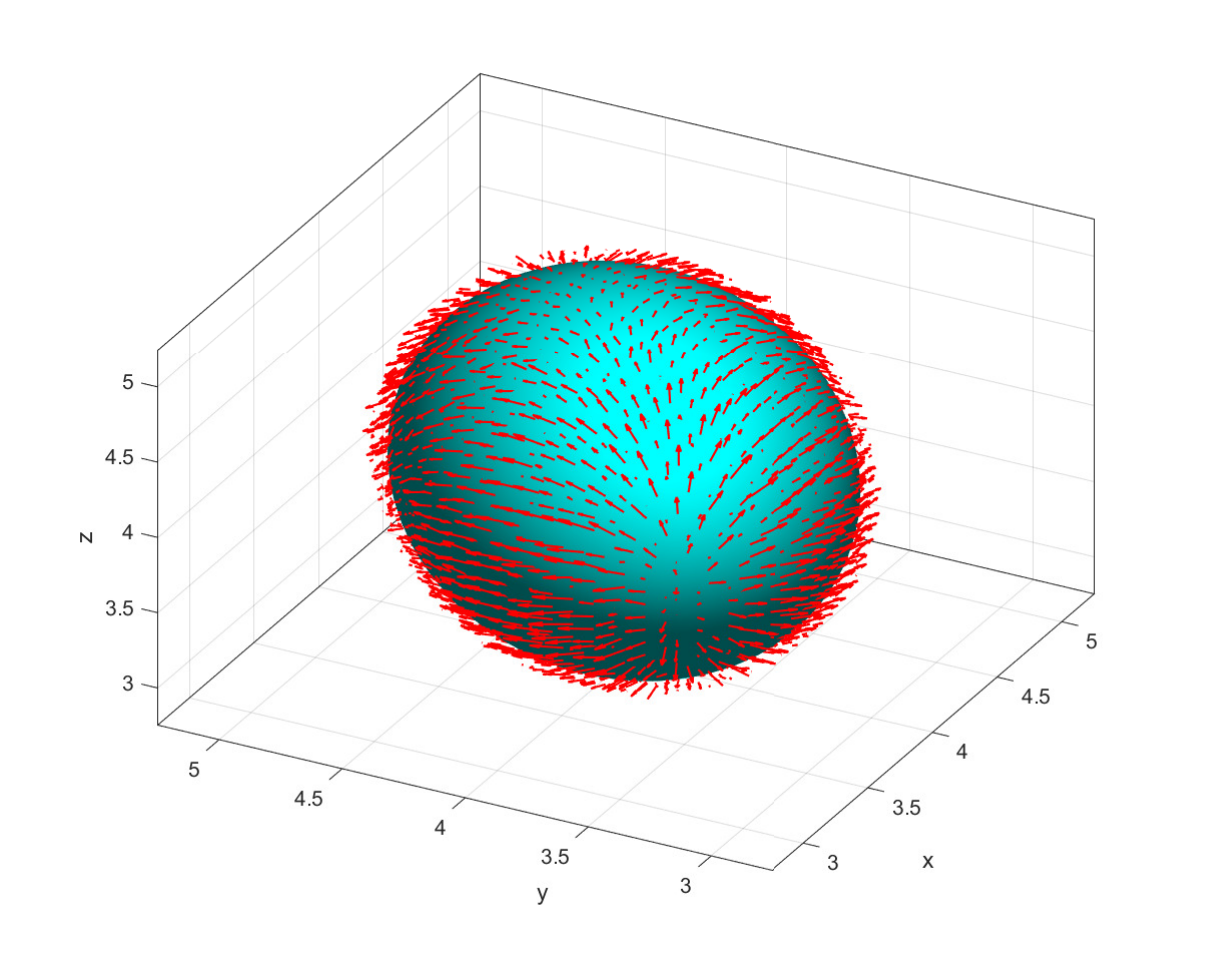} 
			\label{shape3ddrop}
		\end{minipage}
	}
	\\
	\centering
	\subfigure[Concentration of surfactants]{
		\begin{minipage}[b]{0.45\textwidth}
			\includegraphics[width=1\textwidth]{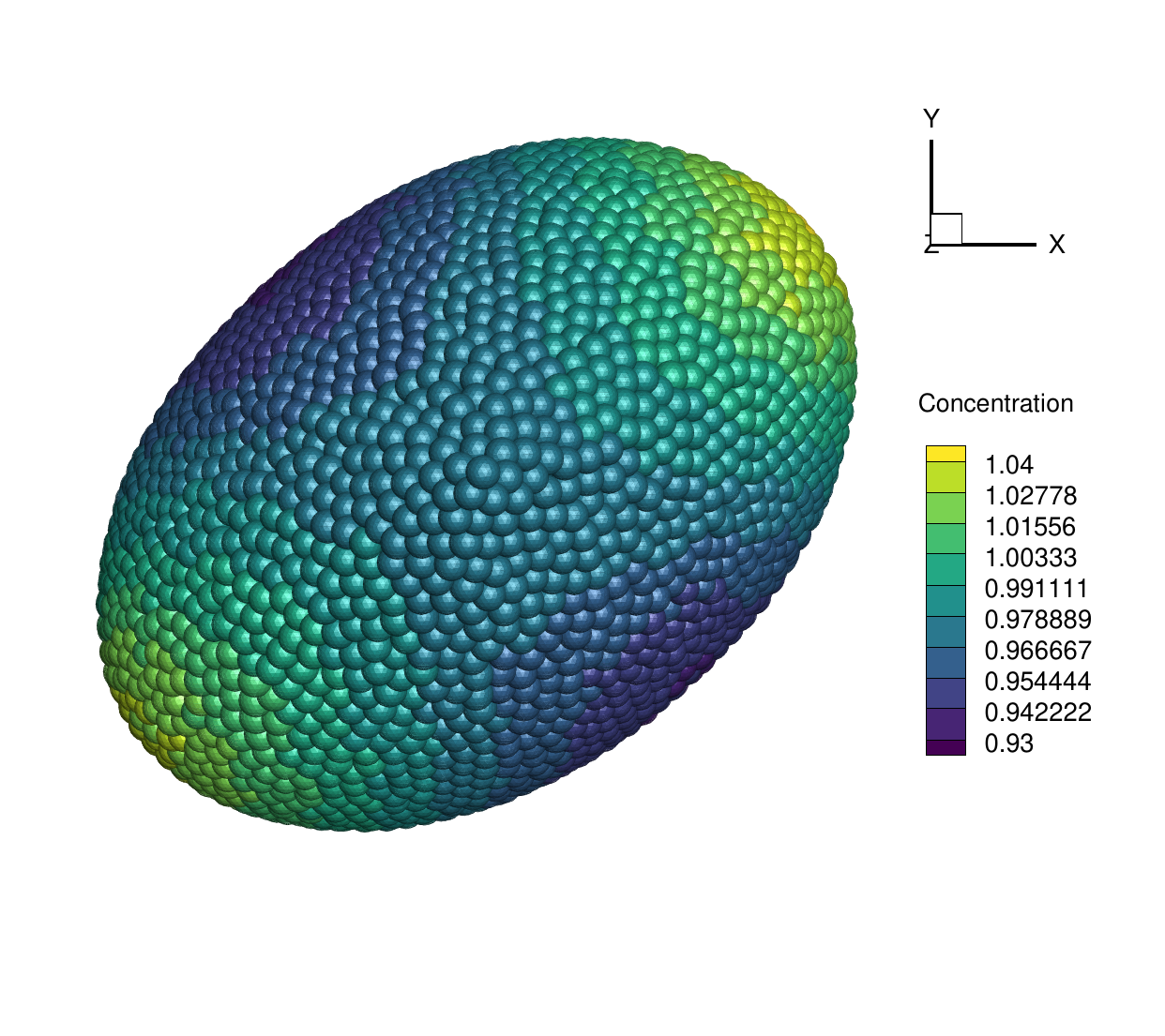} 
			\label{concen3ddrop}
		\end{minipage}
	}
	\caption{Droplet deformation with Marangoni force at $t=8$, with a grid resolution of $96\times96\times96$.}
\end{figure}

Droplet deformation leads to a higher surfactant concentration near the tips and a lower concentration elsewhere. This phenomenon results in reduced surface tension near the tips and increased tension elsewhere, with a deformation degree of $\mathcal{D}_{s}=0.1954$. %larger than the clean interface $D_c = 0.1899$.
When also considering the Marangoni force induced by the concentration gradient, as shown in Fig. \ref{concen3ddrop}, the Marangoni force acting in the direction opposite to the concentration gradient, as shown in Fig. \ref{shape3ddrop}, decreases the deformation, with $\mathcal{D}_{Ma}=0.1888<\mathcal{D}_{s}=0.1954$.

%{\color{red}Moreover, the remeshing process contributes to only $1.5\%$ of the total computational expenses, demonstrating the good efficiency of the present remeshing method in 3-D.}

\section{Conclusion}
\label{conclusion}
In this paper, we develop a generalized hybrid method for surfactant dynamics. To prevent particle clustering, a remeshing method is employed. By optimizing the mass redistribution during remeshing,  the present hybrid method is applicable for both 2-D and 3-D cases. 
Through testing the translation of circle/sphere, and deformation of a circle/sphere in a given shear flow, we find that present method has comparable accuracy and convergence rate to those reported by Fan et.al \cite{fan2023hybrid}. Additionally, the dumbbell case shows that the present method has better robustness, which can handle the topology change without any extra treatments. This is attributed to our method's ability to optimally select redistribution weights with least weighted squares, thereby avoid the occurrence of weights with large magnitudes. The case of 2-D droplet deformation with surfactant dynamics shows that the present method also has comparable conservation property and efficiency to the findings in Fan et al \cite{fan2023hybrid}. We validate the present method for 3-D cases, which confirms its consistent accuracy, convergence, and effectiveness in a higher-dimensional applications. However, this field remains unsolved questions for further exploration, regarding the enhancement of accuracy and more effective implementation strategies for parallelization in distributed-memory architectures.

%% The Appendices part is started with the command 
\appendix
\section{Derivation of analytical solution of given shear flow in 3-D}
\label{derivation}
Consider a sphere with radius $r_0$ deformed in a given shear flow: $u=w=0$, $v=Mx$, where $M=\partial v/\partial x$ is the shear rate.
For an arbitrary piece of surface $S(t)$ at time instant $t$, the governing equation of concentration on this surface is:
\begin{equation}
	\frac{d}{dt}\int_{S(t)}{C(\theta,\phi,t)dS(t)}=0,
	\label{csv_shear}
\end{equation}
where $(r,\theta,\phi)$ is the spherical polar coordinates and the origin is at the center of the given sphere. For simplicity, we use $dS(t)$ as the abbreviation of $dS(\theta,\phi,t)$.
Integrating Eq. (\ref{csv_shear}), we have the mass conservation for $\forall t \geq 0$
\begin{equation}
	\int_{S(t)}{C(\theta,\phi,t)dS(t)}=\int_{S(0)}{C(\theta,\phi,0)dS(0)}.
	\label{eq:conservshear}
\end{equation}

The key is to transforming $dS$ as $T(\theta,\phi,t)d\theta d\phi$, where coefficient $T$ can be calculated by comparing 
\begin{equation} 
	dS(t) = \Vert \frac{\partial \mathbf{r}(t)}{\partial \phi} \times \frac{\partial \mathbf{r}(t)}{\partial \theta} \Vert d\theta d\phi= T(\theta,\phi,t)d\theta d\phi,
	\label{transformt}
\end{equation}
and
\begin{equation} 
	dS(0) = \Vert \frac{\partial \mathbf{r}(0)}{\partial \phi} \times \frac{\partial \mathbf{r}(0)}{\partial \theta} \Vert d\theta d\phi=T(\theta,\phi,0) d\theta d\phi.
	\label{transform0}
\end{equation}
Here $\mathbf{r}(t)$ is the corresponding position vector at time $t$.
Substituting $dS(t)$ and $dS(0)$ into Eq. (\ref{eq:conservshear}) , we derive
\begin{equation}
	\int_{S(t)}{C(\theta,\phi,t)T(\theta,\phi,t)d\theta d\phi}=\int_{S(0)}{C(\theta,\phi,0)T(\theta,\phi,0)d\theta d\phi},
\end{equation}
Because the element $dS(t)$ is an arbitrary element, the concentration field always satisfies
\begin{equation}
	C(\theta,\phi,t)=\frac{C(\theta,\phi,0)T(\theta,\phi,0)}{T(\theta,\phi,t)}=\frac{T(\theta,\phi,0)}{T(\theta,\phi,t)}.
	\label{eq:exactconcentration}
\end{equation}

From initial condition, we have
\begin{equation}
	\mathbf{r}(0)=\left [
	\begin{aligned}
		&r_0\sin(\phi)\cos(\theta)\\
		&r_0\sin(\phi)\sin(\theta)\\
		&\quad \  r_0\cos(\phi)
	\end{aligned}
	\right ],
	\theta \in [0,2\pi] \ \phi \in [0,\pi].
\end{equation}
On the other side, according to the velocity field, the position vector $\mathbf{r}(t)$ can be obtained as
\begin{equation}
	\mathbf{r}(t)=\left [
	\begin{aligned}
		&\quad \quad \quad \quad \quad r_0\sin(\phi)\cos(\theta)\\
		&r_0\sin(\phi)\sin(\theta)+Mr_0t\sin(\phi)\cos(\theta)\\
		&\quad \quad\quad \quad \quad \quad \  r_0\cos(\phi)
	\end{aligned}
	\right ].
\end{equation}
Calculating the partial derivatives of $\mathbf{r}(t)$ and $\mathbf{r}(0)$, the coefficient $T(\theta,\phi,t)$ and $T(\theta,\phi,0)$ can be derived by Eqs. (\ref{transformt}) and Eq. (\ref{transform0}), respectively.
Finally substituting $T(\theta,\phi,t)$ and $T(\theta,\phi,0)$ into Eq. (\ref{eq:exactconcentration}), we obtain the exact solution:
\begin{equation}
	C(\phi,\theta,t)=\frac{1}{\sqrt{M^2t^2\sin^2(\phi)\sin^2(\theta)-2Mt\sin^2(\phi)\sin(\theta)\cos(\theta)+1}}.
\end{equation}



\bibliographystyle{unsrt}
\bibliography{references}

\begin{thebibliography}{10}

\bibitem{gaver1990dynamics}
Donald~P Gaver and James~B Grotberg.
\newblock The dynamics of a localized surfactant on a thin film.
\newblock {\em Journal of Fluid Mechanics}, 213:127--148, 1990.

\bibitem{adami2010conservative}
Stefan Adami, Xiangyu Hu, and Nikolaus~A Adams.
\newblock A conservative sph method for surfactant dynamics.
\newblock {\em Journal of Computational Physics}, 229(5):1909--1926, 2010.

\bibitem{hu2007incompressible}
Xiangyu Hu and Nikolaus~A Adams.
\newblock An incompressible multi-phase sph method.
\newblock {\em Journal of computational physics}, 227(1):264--278, 2007.

\bibitem{xu2012dynamic}
JH~Xu, PF~Dong, H~Zhao, CP~Tostado, and GS~Luo.
\newblock The dynamic effects of surfactants on droplet formation in coaxial
  microfluidic devices.
\newblock {\em Langmuir}, 28(25):9250--9258, 2012.

\bibitem{xu2003eulerian}
Jian-Jun Xu and Hong-Kai Zhao.
\newblock An eulerian formulation for solving partial differential equations
  along a moving interface.
\newblock {\em Journal of Scientific Computing}, 19(1):573--594, 2003.

\bibitem{adalsteinsson2003transport}
David Adalsteinsson and James~A Sethian.
\newblock Transport and diffusion of material quantities on propagating
  interfaces via level set methods.
\newblock {\em Journal of Computational Physics}, 185(1):271--288, 2003.

\bibitem{xu2006level}
Jian-Jun Xu, Zhilin Li, John Lowengrub, and Hongkai Zhao.
\newblock A level-set method for interfacial flows with surfactant.
\newblock {\em Journal of Computational Physics}, 212(2):590--616, 2006.

\bibitem{SCHRANNER2016653}
Felix~S. Schranner and Nikolaus~A. Adams.
\newblock A conservative interface-interaction model with insoluble surfactant.
\newblock {\em Journal of Computational Physics}, 327:653--677, 2016.

\bibitem{teigen2011diffuse}
Knut~Erik Teigen, Peng Song, John Lowengrub, and Axel Voigt.
\newblock A diffuse-interface method for two-phase flows with soluble
  surfactants.
\newblock {\em Journal of computational physics}, 230(2):375--393, 2011.

\bibitem{olshanskii2014stabilized}
Maxim~A Olshanskii, Arnold Reusken, and Xianmin Xu.
\newblock A stabilized finite element method for advection--diffusion equations
  on surfaces.
\newblock {\em IMA journal of numerical analysis}, 34(2):732--758, 2014.

\bibitem{james2004surfactant}
Ashley~J James and John Lowengrub.
\newblock A surfactant-conserving volume-of-fluid method for interfacial flows
  with insoluble surfactant.
\newblock {\em Journal of computational physics}, 201(2):685--722, 2004.

\bibitem{wang2021thin}
Mengdi Wang, Yitong Deng, Xiangxin Kong, Aditya~H Prasad, Shiying Xiong, and
  Bo~Zhu.
\newblock Thin-film smoothed particle hydrodynamics fluid.
\newblock {\em ACM Transactions on Graphics (TOG)}, 40(4):1--16, 2021.

\bibitem{hou1994removing}
Thomas~Y Hou, John~S Lowengrub, and Michael~J Shelley.
\newblock Removing the stiffness from interfacial flows with surface tension.
\newblock {\em Journal of Computational Physics}, 114(2):312--338, 1994.

\bibitem{lai2008immersed}
Ming-Chih Lai, Yu-Hau Tseng, and Huaxiong Huang.
\newblock An immersed boundary method for interfacial flows with insoluble
  surfactant.
\newblock {\em Journal of Computational Physics}, 227(15):7279--7293, 2008.

\bibitem{botsch2004remeshing}
Mario Botsch and Leif Kobbelt.
\newblock A remeshing approach to multiresolution modeling.
\newblock In {\em Proceedings of the 2004 Eurographics/ACM SIGGRAPH symposium
  on Geometry processing}, pages 185--192, 2004.

\bibitem{fan2023hybrid}
Yu~Fan, Yujie Zhu, Xiaoliang Li, Xiangyu Hu, and Nikolaus~A Adams.
\newblock A 2d hybrid method for interfacial transport of passive scalars.
\newblock {\em arXiv preprint arXiv:2304.09550}, 2023.

\bibitem{davis1975interpolation}
Philip~J Davis.
\newblock {\em Interpolation and approximation}.
\newblock Courier Corporation, 1975.

\bibitem{zhu2021cad}
Yujie Zhu, Chi Zhang, Yongchuan Yu, and Xiangyu Hu.
\newblock A cad-compatible body-fitted particle generator for arbitrarily
  complex geometry and its application to wave-structure interaction.
\newblock {\em Journal of Hydrodynamics}, 33(2):195--206, 2021.

\bibitem{levin1998approximation}
David Levin.
\newblock The approximation power of moving least-squares.
\newblock {\em Mathematics of computation}, 67(224):1517--1531, 1998.

\bibitem{luo2015conservative}
Jian Luo, Xiangyu Hu, and Nikolaus~A Adams.
\newblock A conservative sharp interface method for incompressible multiphase
  flows.
\newblock {\em Journal of Computational Physics}, 284:547--565, 2015.

\bibitem{taylor1934formation}
Geoffrey~Ingram Taylor.
\newblock The formation of emulsions in definable fields of flow.
\newblock {\em Proceedings of the Royal Society of London. Series A, containing
  papers of a mathematical and physical character}, 146(858):501--523, 1934.

\bibitem{taylor1932viscosity}
Geoffrey~Ingram Taylor.
\newblock The viscosity of a fluid containing small drops of another fluid.
\newblock {\em Proceedings of the Royal Society of London. Series A, Containing
  Papers of a Mathematical and Physical Character}, 138(834):41--48, 1932.

\end{thebibliography}
\end{document}